\documentclass[12pt,preprint]{aastex}

\newcommand{\tlusty}{{\sc Tlusty}}
\newcommand{\synspec}{{\sc Synspec}}
\newcommand{\teff}{$T_{\rm eff}$}

\newcommand{\Mdot}{\hbox{$\dot M$}}
\newcommand{\msolyr}{M$_{\odot}$\,yr$^{-1}$}
\newcommand{\fuse}{{\sl FUSE\/}}

\slugcomment{to appear in {\sl Astrophys. J.\/}, February 2004}

\shorttitle{FUSE Observations of He-rich sdB Stars}
\shortauthors{Lanz et al.}

\begin{document}

\title{Flash Mixing on the White Dwarf Cooling Curve: \\
          FUSE Observations of three He-rich sdB Stars\altaffilmark{1}}

\altaffiltext{1}{Based on observations made with the NASA-CNES-CSA
      Far Ultraviolet Spectroscopic Explorer. \fuse\  is operated for NASA
      by the Johns Hopkins University under NASA contract NAS5-32985.}

\author{Thierry Lanz\altaffilmark{2,3}, Thomas M. Brown\altaffilmark{4},
           Allen V. Sweigart\altaffilmark{3}, \\
           Ivan Hubeny\altaffilmark{3,5}, and Wayne B. Landsman\altaffilmark{3,6}}

\altaffiltext{2}{Department of Astronomy, University of Maryland,
College Park, MD 20742}
\altaffiltext{3}{NASA Goddard Space Flight Center, Code 681, Greenbelt, MD 20771}
\altaffiltext{4}{Space Telescope Science Institute, 3700 San Martin
Drive, Baltimore, MD 21218}
\altaffiltext{5}{National Optical Astronomy Observatory, Tucson, AZ 85726}
\altaffiltext{6}{Science Systems \& Applications, Inc.}

\email{tlanz@umd.edu, tbrown@stsci.edu, sweigart@bach.gsfc.nasa.gov,
         hubeny@noao.edu, landsman@mpb.gsfc.nasa.gov}

\begin{abstract}
We present {\sl Far-Ultraviolet Spectroscopic Explorer\/}
(\fuse) spectra of three He-rich sdB stars. Two of these
stars, PG1544+488 and JL87, reveal extremely strong \ion{C}{3} lines
at 977 and 1176\,\AA, while the carbon lines are quite weak in the
third star, LB1766. We have analyzed the \fuse\  data using \tlusty\
NLTE line-blanketed model atmospheres, and find that PG1544+488 has a
surface composition of 96\%~He, 2\%~C, and 1\%~N.
JL87 shows a similar surface
enrichment of carbon and nitrogen, but some significant
fraction of hydrogen still remains in its atmosphere. Finally,
LB1766 has a surface composition devoid of hydrogen and
strongly depleted of carbon, indicating that its surface
material has undergone CN-cycle processing.

We interpret these observations with new evolutionary
calculations which suggest that He-rich sdB stars with C-rich
compositions are the progeny of stars which underwent
a delayed helium-core flash on the
white-dwarf cooling curve. During such a flash the interior
convection zone will penetrate into the hydrogen envelope,
thereby mixing the envelope with the He- and C-rich core.  Such
``flash-mixed'' stars will arrive on the extreme horizontal branch
(EHB) with He- and C-rich surface compositions and will be hotter than
the hottest canonical (i.e., unmixed) EHB stars.  Two types of flash
mixing are possible: ``deep'' and ``shallow'', depending on whether the
hydrogen envelope is mixed deeply into the site of the helium
flash or only with the outer layers of the core.  Based
on both their stellar parameters
and surface compositions, we suggest that PG1544+488 and JL87 are
examples of ``deep'' and ``shallow'' flash mixing, respectively.
Flash mixing may therefore represent a new evolutionary channel
for producing the hottest EHB stars.
However, flash mixing cannot explain the abundance pattern
in LB1766, which remains a challenge to current evolutionary models.
\end{abstract}



\section{Introduction}
\label{Intro}

The formation mechanisms of subdwarf B (sdB) stars have been the source of
considerable interest for decades. In the Galactic field, sdB stars are
distinguished by their high effective temperatures, $T_{\rm eff} > 25\,000$\,K, and
high surface gravities, $\log g > 5$. In the globular clusters they are identified
with the extreme horizontal branch (EHB) stars found at the faint end of the long
blue HB tail in such clusters as NGC~6752 and M~13. The range in
$T_{\rm eff}$ on the HB comes from the range in envelope mass that  
lies above the hot helium-burning core. Stars at the cool end of the HB have
a large envelope ($\approx 0.2 - 0.4\,M_\odot$), while stars at the hot end of
the HB have an extremely thin envelope ($\approx 10^{-3}\,M_\odot$). This range in
envelope mass is the result of a range in mass loss during the red giant branch (RGB)
phase, where the hottest HB stars have lost several tenths of a solar mass.

Normally, low-mass stars ignite helium at the tip of the RGB during the
helium-core flash. However, if
they undergo very high mass loss, stars will leave the RGB and ignite
helium while
descending the white-dwarf (WD) cooling curve \citep{cas293}.
\cite{dcruz96} suggested that such stars might provide another avenue
for populating the hot end of the HB. When stars
undergo a late helium-core flash on the WD cooling curve, \cite{Sweigart97}
showed that the thin hydrogen envelope will be mixed with the helium- and
carbon-rich core. The high helium-burning luminosity at the peak of
the flash ($\approx 10^{10}~L_\odot$) drives a temporary convection zone
that extends from the site of the flash outward through the
core.  Ordinarily this flash convection zone does not penetrate into the
hydrogen envelope due to the high entropy of the hydrogen-burning
shell. However, when the flash happens on the WD
cooling curve, it does so in the presence of a much weaker
hydrogen-burning shell.
The associated entropy barrier is then lower and can be overcome,
allowing the flash
convection zone to penetrate into the hydrogen envelope. The thin
hydrogen envelope
is quickly mixed into the helium-burning core. This ``flash mixing''
produces a hydrogen-deficient, helium-burning star with enhanced
helium and carbon
in the envelope. A similar process has been proposed to explain the
extremely hydrogen-poor
R Coronae Borealis stars (also known as ``born-again'' stars), whereby
the surface
hydrogen is mixed inward during a helium-{\em shell} flash on the WD
cooling curve \citep{renzini90}.

Flash-mixed HB stars might explain the anomalies in the hot HB
population of the
globular cluster NGC~2808 (Brown et~al. 2001, hereafter Paper~I;
Sweigart et~al. 2002). Optical color-magnitude
diagrams (CMDs) of this cluster show a gap within the EHB
\citep{sosin97,walker99,bedin00},
while an ultraviolet CMD shows a significant population of stars lying
up to 0.7\,mag below
the canonical EHB (Paper~I).  Neither of these anomalies can be
explained by
canonical HB models. Subluminous EHB stars also appear in the
ultraviolet CMD
of $\omega$~Centauri \citep{dcruz00}. It remains unclear, however, why
we find these subluminous EHB stars in NGC~2808 and in $\omega$~Cen,
but not in other globular clusters with extended blue HB morphologies,
such as NGC~6752 and M~13.

In order to understand the origin of these hot HB anomalies,
\citet{brown2001}
systematically explored the effect of increased mass loss
on the evolutionary track of a low-mass star at the metallicity of
NGC~2808. They found that the overall
properties of the hot HB stars are significantly altered as soon as
flash mixing takes place. As long as the stars maintain a thin
hydrogen-rich envelope, they will be
found at the hot end of the HB at effective temperatures up to $\approx
30\,000$\,K. However, all flash-mixed HB stars with helium-rich
envelopes are hotter by about 5\,000 to 6\,000\,K, creating a gap in
the predicted HB distribution in optical CMDs. Moreover,
model atmosphere calculations show
that the flash-mixed stars are fainter in the
ultraviolet due to flux redistribution between the extreme-UV and the
far-UV. The reduced hydrogen opacity below 912~\AA~permits more
of the flux to be radiated
in the extreme UV at the expense of the flux at longer wavelengths.
For these reasons \citet{brown2001} suggested that both the gap
within the EHB and the subluminous EHB stars in NGC~2808 are the result
of flash mixing on the WD cooling curve.

Extreme HB stars are very faint objects, and consequently the
quantitative spectroscopic analyses
of these stars remain scarce in the literature.
Recently, \citet{moehler02} reported on an analysis of low resolution,
optical spectra
of very hot HB stars in $\omega$~Centauri. Despite the limited quality
of the data,
\citet{moehler02} derived high temperatures and high helium abundances
in general
support of the flash-mixing scenario. On the other hand, \cite{AJ03}
have just completed
an analysis of low resolution optical spectra of 17 helium-rich  
subdwarf (He-sdB) stars in the Galactic field.
In general, they derive low surface gravities ($\log g\approx 5$),
which they argue do not agree with the flash-mixing scenario.
The surface gravities
of the He-sdB stars thus provide an important diagnostic for
deciphering the evolutionary origin of these stars. Quite  
interestingly, about half of the stars in these two samples show
an indication of carbon enrichment
(\ion{C}{2}\,$\lambda$4267), but a quantitative analysis will require
higher quality spectra.

One would also expect to find flash-mixed EHB stars
in the Galactic field. In this case, we need to identify them
spectroscopically.
He-sdB stars are likely candidates.
Subdwarfs are generally classified according to the \ion{He}{1},
\ion{He}{2}, and Balmer lines \citep{PGCat86,saffer94,Class97}. The
optical spectrum
of most sdB stars reveals strong, broad Balmer lines, indicating high
surface gravity,
and little or no \ion{He}{1} absorption. However, about 5\% of the sdB
stars exhibit
strong helium lines, and weak or even no Balmer lines. PG1544+488 is
the prototype
of this spectral class, designated originally as sdO(D) stars by
\cite{PGCat86}.
The true fraction of sdB stars that are helium-rich is unknown, because
surveys of
sdB stars are skewed toward the normal subtype as strong Balmer lines
are often one of the search criteria.

Because field He-sdB stars can be found at shorter distances than the
hot EHB stars in
globular clusters, we can more easily obtain high-resolution spectra at
high signal-to-noise,
appropriate for detailed spectroscopic studies.  In this paper we
report on the
quantitative analysis of the three field He-sdB stars PG1544+488, JL87
and LB1766, using ultraviolet spectra
obtained with the {\sl Far-Ultraviolet Spectroscopic Explorer\/}
(\fuse). Our purpose in selecting these three He-sdB stars
was to test the flash-mixing scenario as a new evolutionary channel for
producing the hottest HB stars. The far-ultraviolet spectral range
is a very rich spectral region, presenting unique opportunities for
deriving the
surface abundances of hot stars. It is particularly well-suited for
confirming the
strong enhancement of helium and carbon with respect to other species.
Sect.~\ref{FUSE} describes the selected
stellar sample and the {\sl FUSE\/} observations. Our non-LTE (NLTE)
line-blanketed model
atmospheres are briefly discussed in Sect.~\ref{NLTE}, while the
results of the
analysis are presented in Sect.~\ref{Analysis}. We report on
new evolutionary calculations for solar composition stars in
Sect.~\ref{Mixing}, and show
that flash mixing can be either deep or shallow, depending on how much
of the core is mixed with the hydrogen envelope.
We discuss our results and their
implication for the formation and evolution of hot HB stars in
Sect.~\ref{ConclSect}.


\section{FUSE Observations}
\label{FUSE}

\subsection{Stellar Sample}
\label{Sample}

The observing program was set up from a list of about 50 He-rich sdB stars
catalogued by \cite{sdb96}. We have selected three relatively bright objects,
with an expected low interstellar extinction, in order to obtain
good S/N ratio spectra with \fuse\ that are little contaminated by
interstellar H$_2$ absorption.

We have selected the class prototype, PG1544+488, as our primary target.
A preliminary analysis that was performed by \cite{Heber88}, and reported
in \cite{HJ97}, indicates that PG1544+488 has an hydrogen-deficient and
helium-rich atmosphere with a significant enhancement of carbon (a mass
fraction of $\approx$1\%).

Our second target is JL87 (=EC~21435-7634), with a blue spectrum revealing
strong \ion{He}{1} lines, Balmer lines (contrary to PG1544+488), and a strong
\ion{C}{2}\,$\lambda$4267 line \citep{Schulz91}. A coarse analysis
by \citet{Schulz91} found a hydrogen abundance $X\approx 0.55$ and
a helium abundance $Y\approx 0.45$, indicating a surface enrichment
in helium. A fine analysis, in particular of the
carbon line, was deferred until better observations become available.
\cite{Kilk95} obtained $UBV$ photometry of JL87, showing mild
interstellar extinction, $E(B-V)\approx 0.1$ mag.
\citet{magee98} measured an apparent rotational velocity,
$V\sin\,i = 120\pm 50$\,km/s, from two optical \ion{He}{1} lines.
JL87 was not part of our original target list proposed for \fuse\
Cycle~3
observations.  However, due to reaction wheel failures in late 2001, it
became difficult to observe targets at low declination.  Consequently
we replaced one of our original
targets, PG1127+019, which lies at a declination of nearly 0$^{\rm o}$,
with JL87, at $-76^{\rm o}$.

Our last program star is LB1766. Low-dispersion Reticon spectra were
obtained by \cite{Kilk92} and show no evidence for hydrogen. Contrary
to the other two stars, these spectra do not suggest strong carbon
enhancement.

\cite{Stark03} have recently revisited the question of binarity among
hot subdwarfs. They have used 2MASS colors from a large sample of
spectroscopically selected sdB and sdO stars to look for
composite colors indicating the presence of late type companions. The
three selected stars have 2MASS colors, $(J-K) < -0.1$, that exclude
the presence of unresolved cool companions which might contribute to
the flux in the \fuse\  spectral range. We cannot rule out the presence
of cooler M-type companions, but any such companions would not affect
the stellar parameters that we derive from the analysis of the \fuse\ 
spectra.

\subsection{Far-Ultraviolet Observations}
\label{Observ}

The three program stars were observed during the prime mission
of \fuse\, between March 2002 and
January 2003, as part of Cycle~3 General Observer Program C129.
Table~\ref{ObsTbl} provides a log of the observations.  PG1544+488 and
JL87 were observed in time-tag mode, while LB1766 was observed in
histogram mode.  Time-tag mode records the position and time of each
photon event on the detector, while histogram mode simply provides the
final integrated exposure on the detector; although the default mode
is time-tag, histogram mode is employed on brighter targets with high
count rates.  All targets were observed through the large $30\arcsec
\times 30\arcsec$ aperture.  The observation of PG1544+488 was
repeated because the side~2 detector was turned off during the first
observation, thus limiting the signal-to-noise ratio and wavelength
coverage.

\subsection{Data Reduction}
\label{DataReduc}

The data were processed with CALFUSE, with slightly different versions,
due to the delay between individual observations and their
analysis. PG1544+488 and JL87 were observed early in the program, and
processed with version 2.1.6; LB1766 was observed much later, and
processed with CALFUSE version 2.2.2.  For each target, we combined
the data from different detector segments and channels to create a
single spectrum, using the IDL fuse\_register code, which was written by
D.~Lindler and provided by the \fuse\  project on
their web site\footnote{http://fuse.pha.jhu.edu}. This
code allows the user to interactively combine data from separate
observations, detector segments, and channels.  Cross-correlation is
used to align the individual spectra, and then a comparison of
overlapping wavelength regions allows the user to find detector
artifacts, which can then be masked before the spectra are combined.
The resulting spectra are shown in Fig.~1.


\section{NLTE Model Atmospheres}
\label{NLTE}

We have analyzed the \fuse\  spectra with NLTE line-blanketed model
atmospheres calculated with our model atmosphere program, \tlusty,
version 200\footnote{Available at http://tlusty.gsfc.nasa.gov}.
\tlusty\  computes stellar model photospheres in a plane-parallel
geometry,
assuming radiative and hydrostatic equilibria. Departures from LTE are
explicitly allowed for a large set of chemical species and arbitrarily
complex model atoms, using our hybrid Complete Linearization/Accelerated
Lambda Iteration method \citep{NLTE1}.

\cite{OS02} have recently completed a large grid of hot model stellar
atmospheres, spanning a range of metallicities, from twice the solar
metallicity to  extremely
metal-poor models. Scaled-solar values of metal abundances were assumed,
with the same scale factor for all species heavier than helium, while the
helium abundance was kept at the solar value. However, the \fuse\
spectra readily show that the three program stars have a non-solar  
composition,
with an abundance pattern that notably differs from scaled-solar values.
Therefore, we have calculated several series of NLTE model atmospheres
with increasing sophistication, namely including in subsequent steps
additional
explicit NLTE species. We have computed limited grids of H-He-C, and
H-He-C-N-Si models, with effective temperatures between 25\,000 and
45\,000\,K, $5.0\leq\log g\leq 7.0$, and various helium to hydrogen abundance
ratios. Model atmospheres with different carbon and nitrogen abundances
have been considered.
In the final step, we have calculated fully-blanketed model
atmospheres, including
40 NLTE ions. All model atmospheres have been calculated with model
atoms described in detail by \cite{OS02}. We have only updated the
\ion{C}{3} model atom to incorporate all individual levels provided by the
Opacity Project \citep{IOP95, ADOC14}. The lowest 34 individual levels
are considered explicitly, and all higher levels are grouped into 12
superlevels.
Nickel is not considered as an explicit NLTE species since we found that
it has a negligible effect on the model atmospheres. The final
model atmospheres include 765 NLTE levels and superlevels.


\section{Spectrum Analysis}
\label{Analysis}

\subsection{Methodology}
\label{Method}

The first series of H-He-C NLTE model atmospheres is used to roughly fix
the main stellar parameters: effective temperature, surface gravity,
helium to hydrogen
abundance ratio, and carbon abundance. At this stage, we have adopted
the carbon ionization
balance as the primary \teff\  indicator. There are numerous lines of
\ion{C}{2} and
\ion{C}{3}, and a few \ion{C}{4} lines in the spectral range covered by
{\sl FUSE\/}.
The dominant ionization stage is \ion{C}{3}. We can therefore derive
carbon abundances
from \ion{C}{3} lines, while \ion{C}{2} and \ion{C}{4} lines are quite
sensitive to \teff.
The \ion{He}{2} line wings are sensitive both to \teff\  and $\log g$.
Surface gravity is determined from the \ion{He}{2} lines, with moderate
accuracy ($\pm$0.3\,dex).
Hydrogen Lyman lines are not detected in two of the three stars,
imposing tight upper limits on the hydrogen abundance. For all models,
we have arbitrarily assumed
a microturbulent velocity, $\xi_{\rm t} = 5$\,km\,s$^{-1}$.

Assuming the atmospheric structure from the H-He-C NLTE models
calculated with \tlusty, we have then computed detailed synthetic spectra
with our spectrum synthesis code, \synspec. Hydrogen, helium, and carbon
lines are calculated with NLTE populations, while LTE is assumed for  
the lines of all other species. This step
provides an estimate of the abundances of other elements, and an
additional check of the effective temperature using
the nitrogen and silicon ionization balances.
We found that \ion{Si}{3}\,$\lambda\lambda$1108-10-13 and
\ion{Si}{4}\,$\lambda\lambda$1122-28
are particularly useful indicators. Although the high surface gravity
of the program stars
ensure that LTE is a reasonable approximation for minor species,
we have calculated additional NLTE models including nitrogen and silicon
as explicit NLTE species. Because a major NLTE effect consists in
shifting ionization balances toward higher ionization relative to the
LTE values, it is essential to account for such possible shifts
when using ionization balances as temperature indicators.

The method that we followed provides a quick path to derive stellar
parameters and surface
abundances of program stars with abundance patterns that are very
different to scaled-solar
abundances. Fully line-blanketed NLTE model atmospheres are calculated
in a final
step. The final fits are presented in the following sections. The
theoretical spectra
are scaled to the {\sl FUSE\/} spectra to match the observed flux
levels around 1050 to
1100\,\AA. The derived normalization factor is then applied to the
theoretical spectrum
throughout the spectral range, without any adjustment. The agreement
between the theoretical
and observed spectra confirms that the three
stars suffer from little differential interstellar extinction
across the \fuse\  bandpass.

A number of strong interstellar lines are present in the {\sl FUSE\/}
spectral range, and
are seen in our data. In some situations it is important to include
these lines in the analysis
in order to achieve a good match to the data, for example in case of
strong blends of photospheric
and interstellar lines. We have derived interstellar column densities
towards the three stars for
a number of ions. However, we have made no attempt at deriving  
consistent
excitation/ionization
models to determine properties of these interstellar lines of sight.
The interstellar lines are
modeled with Voigt profiles. In addition to atomic lines, we have also
accounted for
the Lyman and Werner band systems of molecular hydrogen. The necessary
H$_2$ data have been
extracted from \cite{kurucz15}. We have determined the H$_2$ column
density by fitting lines
from the ground state. Excitation temperatures are empirically
determined to match
H$_2$ lines from excited levels.

To achieve a fit to the data, the following four additional steps need
to be applied to the \synspec-generated spectra:
{\sl (i)} rotational convolution; {\sl (ii)} interstellar atomic and
molecular
line absorption; {\sl (iii)} instrumental convolution; and {\sl (iv)}
flux normalization.

\subsection{PG1544+488}
\label{PG1544}

The {\sl FUSE\/} spectrum of the archetype He-sdB star, PG1544+488,
reveals striking features,
most notably very strong \ion{C}{3}\,$\lambda$977, $\lambda$1176 lines,
and the absence of
photospheric hydrogen Lyman lines (only narrow interstellar Lyman lines
are present). The
spectrum looks thus quite abnormal around Ly\,$\gamma$ and
\ion{C}{3}\,$\lambda$977.

Therefore, we have started the analysis of PG1544+488 assuming an
atmosphere
almost devoid of hydrogen, adopting a low hydrogen mass fraction,
$X=0.07\%$
\citep{Heber88}. We have constructed a small grid of H-He-C model
atmospheres, spanning
a range of effective temperatures between 30\,000 and 40\,000\,K,
surface gravity
($5.6\leq\log g\leq 6.4$), and carbon abundance (mass fraction from 1\%
to 5\%).
The best match was obtained for \teff = 36\,000\,K, $\log g=6.0$, and a
carbon
mass fraction of 2\%. Most carbon lines are saturated; however, carbon
mass fractions
as low as 1\% or as high as 3\% can be excluded by our analysis.
\ion{C}{2} lines (e.g., $\lambda$1092) tend to favor a lower effective
temperature (34\,kK), while \ion{C}{2}\,$\lambda$1010, 1066, are best
fitted with
a higher temperature (38\,kK) model. The \ion{C}{4} lines
($\lambda$1107, 1169) favor
\teff$\approx 36\,000$\,K.

As outlined in the previous section, we have added N and Si as explicit
NLTE species
in the next step, and included all other species in LTE.
The nitrogen and silicon ionization balances support the effective
temperature
derived from carbon lines. \ion{P}{4} and \ion{P}{5} resonance lines
are also well matched,
providing further support to the adopted effective temperature, \teff\
= 36\,000\,K.
We estimate the uncertainty on \teff\ to be $\pm$2\,000\,K. The {\sl
FUSE\/} data and
our analysis cannot be reconciled with the lower effective temperature
(31\,kK) derived
by \citet{Heber88}. The effective temperature derived by \citet{AJ03},
34\,000$\pm$300\,K,
agrees better, although the quoted uncertainty seems significantly too
low.

\ion{He}{2}\,$\lambda$1085 line wings are best matched assuming a
surface gravity $\log g=6.0$.
We have used the approximate, but sufficiently accurate, formalism of
\cite{HHL94} to
model the line profiles of the higher members of the Balmer series of
\ion{He}{2}.
This approach provides a smooth transition between a Doppler core and a
quasistatic Holtzmark wing. Although the
{\sl FUSE\/} spectral range does not show features that are very
sensitive to gravity,
we may exclude surface gravities lower than $\log g\leq 5.5$ or higher
than $\log g\geq 6.5$. The derived gravity may be checked for consistency
against predictions from stellar structure models. Hot HB stars have
indeed well
defined masses, $M\approx 0.5\,M_\odot$, and a limited range in
luminosities,
$1.1\leq\log L/L_\odot\leq 1.5$. Adopting \teff\ = 36000$\pm$2000\,K
and $\log g = 6.0\pm0.3$,
we derive a luminosity for PG1544+488, $\log L/L_\odot = 1.3\pm0.4$. We
cannot reconcile
our result with the much lower gravity derived by \citet{Heber88} and
\citet{AJ03},
$\log g = 5.1\pm0.1$. \citet{AJ03} derived systematically low gravities
($\log g\approx 5$) for all He-rich sdB stars.
Since different scenarios of the origin of He-rich sdB stars would be
implied by these
different values of the surface gravity, we need to understand the
reason behind this discrepancy. We will address this question in
Sect.~\ref{ConclSect}.

Besides helium and carbon, nitrogen is also significantly enhanced. We
have derived
a nitrogen mass fraction of 1\%. Silicon and sulfur lines may have
slightly subsolar
surface abundances, while \ion{P}{4} and \ion{P}{5} resonance lines
yield a phosphorus
surface enrichment by about a factor 3 compared to the solar phosphorus
abundance. The
Si-P-S abundance pattern may be an indication that radiative levitation
and gravitational
settling may be affecting the surface composition, though the evidence
is not strong.
The numerous weak Fe lines are well matched assuming a solar
metallicity. Finally,
we can set an upper limit to the hydrogen photospheric mass fraction,
$X<0.002$, from
the absence of Lyman line wings. This limit is consistent with the
hydrogen mass
fraction derived from optical spectroscopy by \citet{Heber88}.

Our results are summarized in Table~\ref{ResuTbl}. Fig.~1 shows the
{\sl FUSE\/} spectra and best model fit for the three program stars.
Generally, the
agreement between the observed and model spectra of PG1544+488 is
excellent. Two
problems may nevertheless be noted. The \ion{C}{3}\,$\lambda$977 line
wings are
too weak in the model spectrum. The Stark broadening parameter would
need to be
increased considerably in order to match the observations.
Additionally, there are
several broad observed features that are absent in the model.
An instrumental origin does not seem plausible,
because the same features appear in spectra from independent detector
segments that
overlap in wavelength coverage at each given feature.
These broad features are probably of photospheric origin, but we have
been unable
to secure an identification. We note that most of them are also present
in JL87
(e.g., $\lambda\lambda$1030, 1060, 1115, 1137), albeit weaker. A broad
absorption
at $\lambda$987-88 may be modeled by several H$_2$ lines in JL87, but
there are no
other indications of molecular hydrogen absorption in PG1544+488
spectrum.

\subsection{JL87}
\label{JL87}

The most striking difference between the PG1544+488 and JL87 spectra is  
the presence of broad hydrogen Lyman lines in the latter, which are most  
obvious near the series limit at wavelengths shorter than 950\,\AA~(see Fig.~1).
The presence of a significant
amount of hydrogen in the photosphere of JL87 is not surprising in view
of the detection of hydrogen Balmer lines reported by \citet{Schulz91}.

From Str\"omgren photometry and Balmer lines, \citet{Schulz91} derived
JL87 stellar parameters, \teff = 28000$\pm$1000\,K,
and $\log g = 5.2\pm 0.3$. Our analysis
of the {\sl FUSE\/} spectrum provides consistent values,
\teff = 29000$\pm$2000\,K, and $\log g = 5.5\pm 0.3$. An effective
temperature lower
than 30000\,K is favored by the strength of
\ion{Si}{3}\,$\lambda$1108-10-13 and
\ion{Fe}{3}\,$\lambda$1088. Model spectra calculated with the two sets
of parameters
are very similar, and we cannot really favor one or the other. We have
adopted the higher values (29\,kK; 5.5) which yield a luminosity, $\log
L/L_\odot = 1.44\pm0.3$, in better agreement with stellar structure
models of hot HB stars.

Optical \ion{He}{1} lines indicate a moderate surface enrichment of
helium, He/H = 0.20$\pm$0.05, by number \citep{Schulz91}. The \fuse\
spectroscopy is consistent with this, but we cannot discriminate between
the solar abundance (He/H = 0.1) and the Schulz et al. value.
We find a significant surface enrichment of carbon and nitrogen (both
by a factor 5), corresponding to mass fractions of 1.4\% and 0.4\%,
respectively. From silicon and iron lines, we find that a solar
metallicity results in a good fit to JL87 spectrum.

Moderately strong absorption by interstellar molecular hydrogen is
apparent in JL87.
A good match of the best model spectrum to the {\sl FUSE\/} data
requires accounting
for H$_2$. We have identified all resonance lines (ground vibrational
level, ground
rotational level $J=0$), most lines from the first
rotational level ($J=1$), and some lines originating from $J=2-4$
levels. To match
the H$_2$ absorption line spectrum, we have adopted a H$_2$ column
density of $1\times 10^{20}$
cm$^{-2}$, and empirical excitation temperatures of 70\,K, 90\,K,
130\,K, and 150\,K, for
$J=1-4$ levels, respectively.

\subsection{LB1766}
\label{LB1766}

LB1766 has been classified as a He-sdB star from low resolution optical
spectroscopy \citep{Kilk92}, with no evidence for hydrogen Balmer lines
at this resolution. However, their spectrum shows \ion{He}{2}\,$\lambda$4686
at about equal strength as \ion{He}{1}\,$\lambda$4713, suggesting higher
temperatures than a typical He-sdB star. In the classifications scheme
of \citet{Class97} the star would be classified as sdO8:He4. Our \fuse\ 
spectrum shows no evidence for Lyman lines, consistent with the absence
of hydrogen in the  optical spectrum. The LB1766 spectrum, however,
presents strong contrasts to the PG1544+488 spectrum, e.g.,
the strong \ion{He}{2}\,$\lambda$972 and weak \ion{C}{3}\,$\lambda$977
in LB1766 (see Fig.~1). LB1766 has a very sharp-lined spectrum,
yielding an upper limit for its apparent rotation,
$V \sin i\leq 25$\,km/s. Similarly to PG1544+488, we do not find
absorption from interstellar H$_2$.

LB1766 has not been previously analyzed quantitatively. We have
determined the effective
temperature from the carbon, \ion{N}{3}/\ion{N}{4}, and
\ion{Si}{3}/\ion{Si}{4}
ionization balances. They yield consistent values, and we have adopted,
\teff = 40000$\pm$2000\,K.
The surface gravity was derived from \ion{He}{2} line wings, and we
find that $\log g$
values in the range between 6.0 and 6.5 result in good fits. We have
adopted
$\log g = 6.3\pm 0.3$. We note that \ion{He}{2}\,$\lambda\lambda$943,
959 are badly
reproduced by the adopted model atmosphere. These two lines are much
stronger in LB1766
than in PG1544+488 (despite similar stellar parameters), and they are
well fitted in the
latter case. From the absence of hydrogen Lyman lines, we derive
an upper limit for the hydrogen abundance, $X\leq 0.0025$, similar to
the case of PG1544+488.

CNO surface abundances depart markedly from the two other stars.
Nitrogen is clearly
overabundant (a factor 7 from solar), while carbon is very deficient
(1/30 of the solar abundance).
\ion{O}{3} lines are clearly identified ($\lambda\lambda$1008, 1040,
1138, 1150-51-54, etc)
and show a moderate oxygen deficiency. The observed CNO abundance
pattern is a signature of material processed through the
CNO-cycle (N/C$>>$1 and N/O$>$1).  Since the CNO-cycle conserves
the total number of CNO nuclei, we would expect the original
CNO abundance in LB1766 to be $\approx 0.009$ by mass, which is
consistent with a metallicity near solar.


\section{Deep and Shallow Flash Mixing}
\label{Mixing}

In Paper~I we explored the different evolutionary tracks which
low-mass stars can follow as they evolve through the helium flash and
showed that
under some circumstances the helium flash will induce substantial
mixing between the helium core and hydrogen envelope. As outlined
in Sect.~\ref{Intro}, our evolutionary sequences with flash
mixing were able to account for a number of anomalies among the
hot HB population in the globular cluster NGC~2808, including the
gap within the EHB and the existence of subluminous EHB stars. These
results raise the possibility that flash mixing might also
be responsible for the formation of the field He-sdB stars. In
this section we will present new evolutionary calculations to
test this possibility. Our goal is to determine if flash-mixed
models can reproduce the surface abundances and stellar
parameters derived in Sect.~\ref{Analysis} for our three He-sdB
stars.

The evolutionary sequences in Paper~I were computed for a metal-poor
composition appropriate for the stars in NGC~2808. Since the
field He-sdB stars are more metal-rich, we have computed a new
set of evolutionary sequences for a solar composition to ensure a
more consistent comparison with the \fuse\ data analysis. Each of
these sequences was evolved continuously from the main sequence
through the helium flash to the zero-age horizontal branch (ZAHB).
The initial main-sequence
mass was $1\,M_\odot$ in all cases. These sequences only
differed in the extent of mass loss along the RGB which we
specified by varying the mass-loss parameter $\eta_{\rm R}$ in
the \citet{reimers75,reimers77} formulation from 0 to 1.3. We emphasize  
that
the Reimers formulation merely provides a numerically
convenient way to vary the amount of mass loss. Whether the actual
mass loss follows a different formulation during single-star
evolution or whether it arises from binary-star interaction
is not important for the present discussion. The important
quantity is the total amount of mass loss on the RGB, not the
process by which the mass is removed.

The different evolutionary tracks followed by our solar
metallicity models during the helium flash are illustrated in
Fig.~2. Normally a low-mass star will ignite helium in its core
at the tip of the RGB and then evolve over $\approx 2\times10^6$ yr
through the helium flash to the ZAHB (Fig.~2a). The
location of a star on the ZAHB depends
on its envelope mass, but as long as the envelope mass is greater
than $\approx 0.01\,M_\odot$, corresponding to a ZAHB
temperature of $\approx 20\,000$\,K, the star will be tightly bound
to the RGB at the time of helium ignition. At higher mass-loss
rates a star will peel off the RGB and evolve to high
effective temperatures before igniting helium,
leading to the so-called ``hot flashers'' \citep[Paper~I]{cas293,
dcruz96}. It is helpful to distinguish between
two types of hot flashers: ``early'' hot flashers, which ignite
helium between the tip of the RGB and the top of the WD
cooling curve (Fig.~2b) and ``late'' hot flashers, which ignite
helium while descending the WD cooling curve (Figs.~2c,d). At
even higher mass-loss rates a star will leave the RGB
and evolve down the WD cooling curve without ever
igniting helium, thus dying as a helium white dwarf.

The high helium-burning luminosity during the helium flash
($\gtrsim 10^9~L_\odot$) produces a temporary
convection zone that extends from the site of the helium flash
outward through the core to just inside the base of the hydrogen
envelope (see Fig. 3). Due to the efficient neutrino cooling of
the central regions of the core, the maximum temperature within
the core, and therefore the site of the helium flash, lie
off-center. Following the flash peak this convection zone retreats
and disappears, although a small convective shell persists in the
outer part of the core for a few thousand years. The carbon
abundance within the flash convection zone ($\approx 0.04$ by
mass) is determined by the amount of helium that a star must burn
in order to lift the degenerate core out of its deep potential
well. In stars which ignite helium either at the tip of the RGB
or as early hot flashers, the flash convection zone does not
penetrate into the hydrogen envelope due to the high entropy
barrier of the hydrogen-burning shell \citep{iben76}. In both
cases the hydrogen-burning shell is a strong energy source
($\approx 2000~L_\odot$) at the onset of the helium flash. Since
the surface composition of such stars is not altered by the
helium flash, we will refer to their evolution as ``canonical'',
as distinguished from the ``noncanonical'' evolution of stars
which undergo flash mixing.

When a star descends the WD cooling curve, the energy
output of its hydrogen shell, and consequently the
entropy barrier of the shell, decrease. As a result, the flash
convection zone is then able to penetrate deeply into the hydrogen
envelope \citep[Paper~I]{Sweigart97}. Hydrogen captured by the
flash convection zone will be mixed into the core while helium and
carbon from the core will be mixed outward into the envelope,
leading to substantial changes in the surface composition. This
flash mixing is a consequence of the basic properties of the
stellar models and should occur whenever a star ignites
helium on the WD cooling curve as a late hot flasher
(Figs.~2c,d). Our calculations have revealed two types of flash
mixing: ``deep'' and ``shallow'', depending on where the flash
occurs along the WD cooling curve. We will discuss these
two types of mixing separately, since they have different
consequences for the surface composition of an EHB star.

Deep mixing occurs when the flash convection zone reaches the hydrogen
envelope shortly after the peak of the flash at the time when the
core is fully convective outside the flash site. In this case,
hydrogen from the envelope will be carried inward towards the hot
helium-burning regions of the core and, along the way, will begin
to react with the abundant $\rm ^{12}C$ nuclei produced by the helium  
burning. The
peak in the hydrogen burning will occur around the layer in the
flash convection zone where the timescale for proton-capture
on $\rm ^{12}C$ is comparable to the timescale for convective mixing.
Thus in the case of deep mixing any hydrogen captured by the flash
convection zone will be burned.

A detailed investigation of the abundance changes produced by
deep mixing poses a major numerical challenge because it requires
one to solve for both the nucleosynthesis and the time-dependent
mixing within the flash convection zone simultaneously. In the
first study of flash mixing, \citet{Sweigart97} was able to
evolve one sequence through the deep mixing phase by ignoring the
energy released by the hydrogen burning. This assumption is
clearly a rough approximation, since, as shown in Paper~I, the
hydrogen burning within the flash convection zone will release a
substantial amount of energy, as much as $\approx$20\% of
the total energy produced during the helium flash. This
additional energy source would be expected to drive the flash
convection zone even further into the hydrogen envelope. Thus the
models of \citet{Sweigart97} should, if anything, underestimate
the extent of the surface composition changes during deep mixing.
Even so, the flash convection zone in these models
captured $\approx$90\% of the envelope hydrogen, resulting in a final
surface composition of 81\% helium and 3\% carbon by mass.

Due to the numerical difficulties in following the composition
changes during deep mixing, \citet{brown2001} stopped their
evolutionary calculations at the onset of mixing and then
restarted them at the ZAHB phase by assuming an envelope
composition of 96\% helium and 4\% carbon by mass. The abundances
of the remaining heavy elements were unchanged from their values
for the adopted metallicity of NGC~2808 (heavy-element abundance
$Z = 0.0015$). \citet{brown2001}
argued that the envelope composition following deep mixing should
reflect the composition of the flash convection zone. We have
adopted the same approach in the present study. The track for the
late hot flasher plotted in Fig.~2d was stopped at the onset of
deep mixing. We then jumped over the flash-mixing phase (dashed
line in Fig.~2d) and constructed a ZAHB model with an envelope
composition of 95\% helium and 4\% carbon, with the remaining
1\% of the envelope consisting of elements heavier than carbon
from the original solar composition.

Recently, \cite{cassisi03} have overcome the numerical
difficulties posed by deep mixing and have followed the detailed
composition changes during this phase. Their calculations confirm
the flash-mixing scenario advocated by \citet{Sweigart97}
and \citet{brown2001}.  In particular, they predict a surface
composition after deep mixing of 96\% helium, 2.9\% carbon
and 0.7\% nitrogen by mass, in good agreement with the surface
composition assumed here. The enhanced nitrogen abundance comes
from the burning of hydrogen on carbon within the flash
convection zone. A tiny fraction of the hydrogen is also expected to
remain in the envelope ($\approx$0.04\% by mass) after deep mixing and  
might be spectroscopically detectable if hydrogen diffuses towards
the surface. We conclude that the surface composition produced by
deep mixing seems well determined. Moreover, it closely resembles
the surface composition of PG1544+488.

In models with deep mixing the full flash convection zone
penetrates into the envelope, and the envelope hydrogen is then mixed
deeply into the core and burned. Such mixing was found during helium
flashes
that began on the WD cooling curve at luminosities
below $\log L/L_\odot \approx 1$. In flashes that began further
up the WD cooling curve, such as the one in Fig.~2c,
only the convective shell that exists in the outer part of the
core once the main body of the flash convection zone recedes (see
Fig.~3) is able to penetrate into the envelope. An example of
this type of mixing, taken from the sequence in Fig. 2c, is shown
in Fig.~4. Unlike the canonical case in Fig.~3, the convective
shell in Fig.~4 moves progressively outward in mass until at time
t = $2\times10^4$ yr after the peak of the flash it makes
contact with the hydrogen envelope. The convective shell
continues to move deeper into the envelope, and at
t = $2.4\times10^4$ yr it joins up with a thin convective
envelope that develops because of the low effective temperatures
of the models during this phase (see Fig.~2c). At this point
helium and carbon from the core are mixed outward to the surface.
We shall refer to this type of flash mixing in which the envelope
is mixed only with the outer convective shell as ``shallow''
mixing.

Shallow mixing differs in one important respect from deep mixing.
Because the temperatures in the outer part of the core are very
low due to the large expansion of the core during the helium
flash, none of the envelope hydrogen captured by the convective
shell will be burned. Thus a star undergoing shallow mixing will
conserve its original hydrogen although this hydrogen will be
diluted with helium- and carbon-rich material from the core. For
the sequence in Fig.~2c the final surface composition after
shallow mixing was 50\% helium, 0.8\% carbon and 0.5\%
nitrogen. The nitrogen abundance is enhanced because the core
material has been processed through the CNO-cycle during the
earlier RGB phase. Because there is no hydrogen burning during
shallow mixing, we were able to evolve our models through this
phase without any of the numerical difficulties encountered
during deep mixing. The final surface composition of the model in
Fig.~2c is similar to the surface composition of JL87, especially
if one allows for a modest enhancement in the surface hydrogen by
diffusion during the HB phase.

Shallow mixing was not found in the low-metallicity sequences
reported in Paper~I. Closer inspection of those sequences showed
that shallow mixing was possible, but only over a much smaller
range in the mass-loss parameter $\eta_{\rm R}$. Although the increment
in $\eta_{\rm R}$ used for the sequences in Paper~I was
very small ($\Delta \eta_{\rm R}$ = 0.001), it was still too large
to detect shallow mixing. The present calculations therefore
suggest that shallow mixing is more likely to occur in
metal-rich stars.

We conclude that deep and shallow flash mixing will produce surface
compositions very similar to those found for PG1544+488 and
JL87. In particular, the robust prediction for the C + N abundance
in the flash-mixed models is nicely confirmed by the
abundances in these two stars.  Flash mixing cannot, however,
explain the very low carbon
abundance of LB1766, which must have formed through
a different evolutionary channel.


\section{Discussion and Conclusions}
\label{ConclSect}

Our results for the three He-sdB stars are summarized in
Table~\ref{ResuTbl} and compared
with EHB evolutionary tracks for canonical, shallow and deep mixing
models
in Fig.~\ref{HRDfig}. All
three stars have an atmosphere rich in helium,
as expected from their He-rich sdB classification, and all show
supersolar
nitrogen abundances. The surface abundance patterns of the three stars,
however,
differ notably, and thus offer a crucial insight into their evolutionary
history. In this section we shall address several points in order to
establish a convincing
scenario of the origin of He-sdB stars. First, we argue that the
surface abundances
cannot be explained by atmospheric processes during the EHB phase, but
rather trace
the nucleosynthetic history of these stars. Second, we address the
reason behind
the discrepancy in the derived surface gravities of the carbon-rich
He-sdB stars
and show that the gravities of these stars are indeed high in accord
with flash-mixing evolutionary tracks.
Finally, we discuss the different abundance patterns and rotational
velocities of the three stars.

A large variety of surface abundances is typical of ``normal'' H-rich
sdB stars, and it
is interpreted as an interplay between diffusion and mass loss. Can we
also explain
our results on He-sdB stars in this atmospheric framework?
\citet{unglaub01} have investigated the effect of weak winds on the
atmosphere of EHB stars.
On timescales short compared to HB lifetimes ($10^8$ years), helium  
settles
down and the
atmosphere becomes very deficient in helium if the wind is very tenuous
(\Mdot\,$\leq 10^{-14}$
\msolyr). Similarly, carbon, nitrogen, and oxygen are also depleted,
although less so,
because their higher electrical charge hampers diffusion. At high mass
loss rates
(\Mdot\,$\geq 10^{-12}$\,\msolyr), the atmospheric chemical composition
remains almost unchanged,
and reflects the envelope composition. The helium deficiencies
typically observed in
sdB stars might be explained by weak winds with \Mdot\,$\approx
10^{-13}$\,\msolyr.
CNO anomalies, including selective nitrogen enrichment, may appear for
special combinations
of mass loss rates and age.
\citet{unglaub01} have also considered
the case of helium-rich atmospheres. They can predict such a He-rich
composition only if a
tiny H-rich envelope ($M\approx 10^{-4}$\,M$_\odot$) is removed by a
strong wind (\Mdot\,$\approx
10^{-12}$\,\msolyr) during the HB phase. At the end of the HB phase, a
He-rich
atmosphere will then appear, with CNO abundances reflecting the
composition of the layers
just inside the base of the hydrogen shell.
Stellar structure calculations indicate that the H-rich envelope is
very likely
much larger than assumed by \citet{unglaub01}, thus requiring a mass
loss rate larger than
$10^{-11}$\,\msolyr\  to remove the hydrogen envelope completely.
Moreover,
mass loss and diffusion do not explain the large surface carbon
abundance.
We therefore conclude that the surface abundances measured in these
three He-sdB stars
cannot be explained by atmospheric processes, and are thus
evidence of {\em nucleosynthetic} processes.

We turn now to the different values of the surface gravity derived for
PG1544+488.
In the previous studies \citep{Heber88, AJ03}, the optical spectrum was
analyzed
with LTE H-He model atmospheres, while we have used ``full'' NLTE model
atmospheres
including 40 NLTE ions in our analysis of the {\sl FUSE\/} spectrum. We
have investigated the
result of using different model atmospheres in the analysis. As a
reference, we adopt the full
NLTE model (\teff=36\,000\,K, $\log g = 6.0$, 2\%~C, 1\%~N).
We have then calculated H-He LTE and NLTE models and H-He-C NLTE models
assuming the same
effective temperature, but different surface gravities
($\log g = 5.0, 6.0$). We have compared
the atmospheric structures and predicted spectra. The H-He model
atmospheres have a much
steeper temperature gradient than our reference NLTE model atmosphere,
while the NLTE H-He-C
models have a temperature structure that is very similar to our
reference model. Fig.~\ref{BlueFig}
illustrates that H-He model atmospheres require lower surface gravity
($\log g\approx 5 - 5.5$)
in order to match the gravity-sensitive wings of the strong optical
\ion{He}{1} lines predicted
by the reference NLTE model ($\log g = 6.0$). With H-He model
atmospheres, we have been unable to
reproduce the optical \ion{He}{2} lines: the strength of
\ion{He}{2}\,$\lambda 4686$ is only
matched with low gravities ($\log g\approx 5$), while the Pickering
series is always predicted
to be too strong  (especially at low gravities). Finally, the shape of
the far-UV continuum is
different, with the flux maximum moving from about 980\,\AA\ (H-He
models) to 1050\,\AA\
in the reference model. Our adopted model atmospheres fit well the
far-UV continuum (see Fig.~1).
Therefore, we conclude that the low surface gravities derived by
\cite{AJ03} are primarily a spurious result of neglecting
carbon (and nitrogen) in the atmospheric structure calculations,
that is, of using H-He model atmospheres. This systematic effect leads
to an underestimate
of $\log g$ by at least 0.5\,dex for carbon-rich He-sdB. Departures
from LTE are only significant
in the core of the strong optical \ion{He}{1} lines, but we could not
assess the resulting
effect on the $\chi^2$ minimization procedure used by \cite{AJ03} to
derive the stellar
parameters.

Our results on PG1544+488, the archetype of He-rich sdB stars, fully
support the flash-mixing
scenario. Its surface composition (H, He, C, and N) matches
the model predictions for deep mixing \citep{brown2001, cassisi03}, and
its high effective temperature and surface gravity coincide nicely
with the EHB evolutionary tracks with deep mixing in Fig.~\ref{HRDfig}.
JL87
also shows an enhancement of carbon and
nitrogen that is similar to PG1544+488, but its atmosphere still
contains a significant amount
of hydrogen. This surface abundance pattern supports our predictions
for shallow flash mixing.
Further making our case is the cooler effective temperature of JL87 near
the hot end of the canonical ZAHB, just as predicted by the
evolutionary tracks with shallow mixing.

A competing theory to explain the origin of hydrogen-deficient evolved
stars (R~CrB stars, extreme helium stars) is the merger of two helium  
white
dwarfs. \cite{saio02} have recently explored the merger of a  
0.6\,$M_\odot$
CO white dwarf with a 0.3\,$M_\odot$ He white dwarf in order to explain
hydrogen-deficient, carbon-rich stars. They are able to reproduce the  
surface
abundance pattern observed in stars like PG1544+488. However, their
model requires fine-tuning in terms of the total mass accreted by
the CO white dwarf and in terms of the mixing needed to bring the  
observed
amounts of carbon and oxygen to the stellar surface. On the other
hand, the flash-mixing model naturally predicts the observed abundances.
The predicted luminosity is however the crucial difference between the
two models: the WD merger scenario implies a post-merger luminosity that
is at least ten times higher than the luminosity predicted by the  
flash-mixing
model. We have argued above that carbon-rich sdB stars have high  
gravities,
hence low luminosities, in accord with the flash-mixing model. A direct
measurement of the luminosity of He-sdB stars will provide  
the decisive test for ruling out one of these two scenarios.

Finally, LB1766 shows a quite different abundance pattern, indicative
of CN-cycle
processing and possibly some limited ON-cycle processing. While CN-cycle
processing
might take place within the H-rich envelope of a star on the RGB, it
would
not produce the very high helium abundance we see today in LB1766. Thus  
the
unanswered question in the case of LB1766 is: What happened to
its hydrogen? We
may envisage two possible scenarios: {\sl (i)\/} the hydrogen was  
burnt; {\sl (ii)\/} the H-rich
envelope was lost. The first possibility seems unlikely. If the hydrogen
was burned via the full CNO-cycle, then the oxygen abundance in
LB1766 should be very small, much less than observed. To be consistent
with the observed oxygen abundance, one would have to argue that the
hydrogen was burned at low temperatures where only the
CN-cycle can operate.  However, it is difficult to think of
an evolutionary phase prior to the EHB where this might
be possible. For example, the ON-cycle operates quite efficiently
at the high temperatures within the hydrogen-burning shell
of a red giant star. As
discussed in Sect.~\ref{Mixing}, the
hydrogen could be burned during a flash-mixing episode, but then we
should see
a high carbon abundance in LB1766, which is not the case. The loss of
the H-rich
envelope during the EHB phase seems equally unlikely. Stellar structure  
models predict a
minimum
envelope mass around $10^{-3}$\,M$_\odot$ at the hot end of the
canonical EHB \citep{dcruz96, brown2001}.
To lose all of this envelope during the EHB phase would require a large
and unlikely mass-loss rate (\Mdot\,$\geq 10^{-11}$\,\msolyr).
Even if all of the hydrogen envelope were removed and the outer layers
of the core exposed, one would still not be able to explain the surface
abundances of LB1766. These layers
were previously processed through the full CNO-cycle during
the RGB phase and would therefore have a very small oxygen abundance.
This is also a problem
if the hydrogen envelope was removed by binary
mass transfer on the RGB.  If even more material were removed
from the core, one would eventually expose layers which were
enriched in carbon by the flash convection zone during the
helium flash. The
derived surface abundance pattern in LB1766, i.e., hydrogen deficiency  
and CNO yields
from the CN-cycle, therefore poses a challenge to current stellar
evolutionary models. It would be
worthwhile to repeat diffusion+wind calculations following
\citet{unglaub01}, assuming
a starting composition given by the flash-mixing scenario, in order to
evaluate the
downward diffusion of carbon.

One unexpected result is the large rotational velocities found for the
carbon-rich
stars PG1544+488 ($V\sin i$ = 100 km/s) and JL87 ($V\sin i$ = 30 km/s).
These values are larger than those typically found for normal sdB stars.
\citet{edel01} estimated  $V\sin i < 5$ km/s for 13 sdB stars observed
at high
signal-to-noise ratio with the FOCES spectrograph,
while \citet{hebe00} found similar limits for
3 sdB stars observed with the Keck HIRES spectrograph. The origin of
this
high rotation rate is not certain. If HB stars have a rapidly rotating
core, as suggested by \citet{sipi00}, then one possibility is that the
flash mixing has brought angular momentum from the core to the surface.
Alternatively, the high rotation rate may reflect the binary origin of
these
system, although in the simplest scenario, where a RGB star loses mass
to a
companion, one expects the angular momentum of the donor to decrease.
We note
that the high rotation may suppress the gravitational settling which
likely
leads to subsolar helium abundances in normal sdB stars. It would
therefore
be of interest to learn if fast rotation is a characteristic shared by
all C-rich sdB stars. Finally, we note that
while the specific angular momentum in PG1544+488 is high compared to
other
field sdB stars, it is several times smaller than in the fastest
rotating blue HB stars observed in globular clusters by \citet{behr03},
which
have $V\sin i = 40$\,km/s, \teff\ = 9000 K, and $\log L = 1.6$.

In conclusion, the observed surface abundance of these three He-sdB
stars clearly
provide evidence of nucleosynthetic processes. Within the error bars,
the derived
stellar parameters and surface composition of PG1544+488 and JL87 agree
with those
predicted for stars with deep and shallow flash mixing, respectively.
However, LB1766
must have had a different evolutionary history, exposing some
remaining puzzles in our understanding of the origin of helium-rich sdB
stars.

\acknowledgments

We thank Michele Stark and Richard Wade for providing 2MASS data ahead
of publication, and Simon Jeffery for discussions on gravity  
determination.
This work was supported by NASA grants NAG5-12383 to the University of
Maryland, and NAG5-12314 to the Space Telescope Science Institute  
(\fuse\
C129 program).



\clearpage

\begin{figure}
\epsscale{0.95}
\plotone{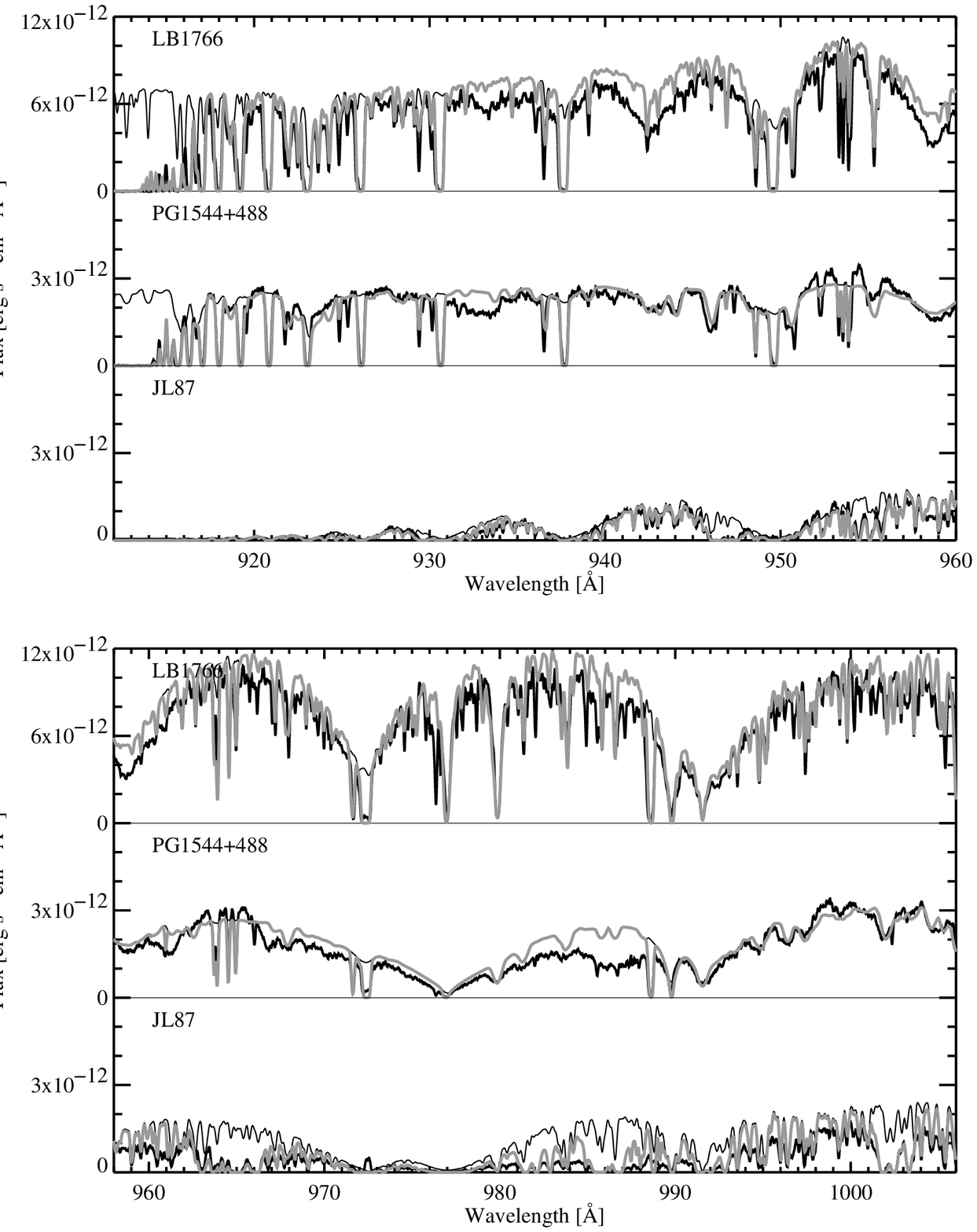}
\figurenum{1a}
\caption{\fuse\  spectra of 3 He-sdB stars (thick black lines) with the
     adopted NLTE model atmosphere fits accounting for interstellar line absorption
     (thick gray line). The thin lines show the NLTE photospheric model spectra.
\label{sdB3Fig}}
\end{figure}

\clearpage

\begin{figure}
\plotone{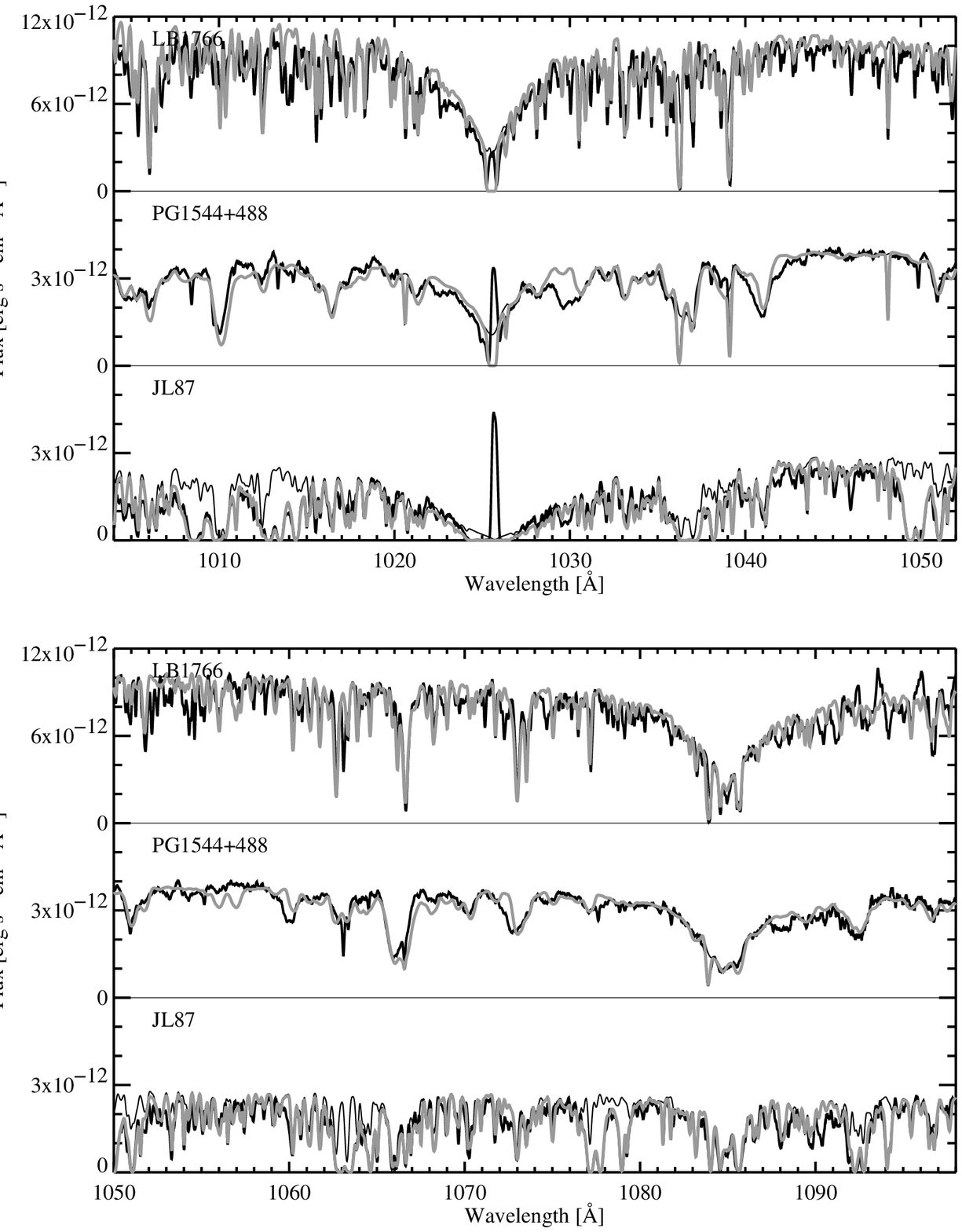}
\figurenum{1b}
\caption{}
\end{figure}

\clearpage

\begin{figure}
\plotone{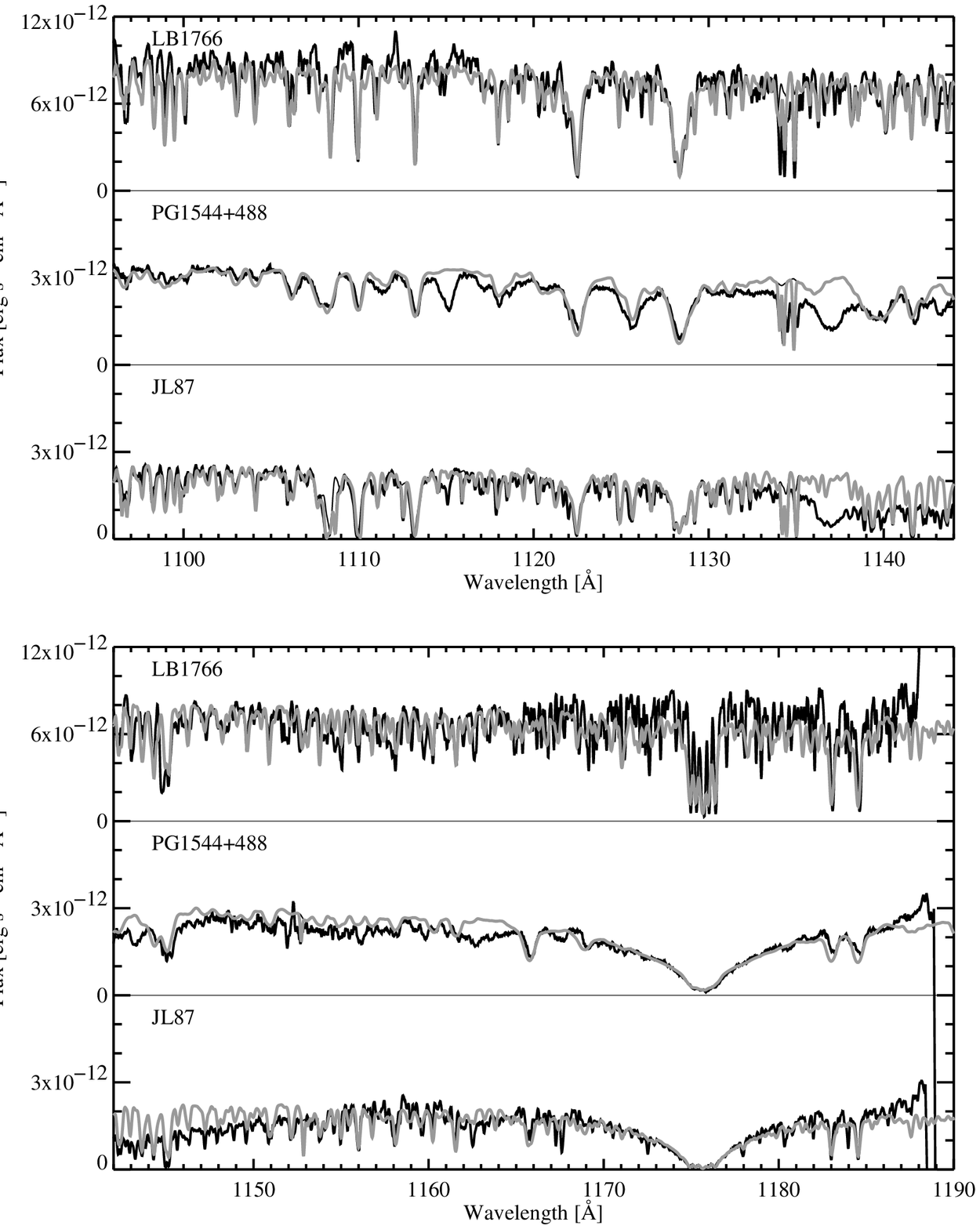}
\figurenum{1c}
\caption{}
\end{figure}
\clearpage

\begin{figure}
\plotone{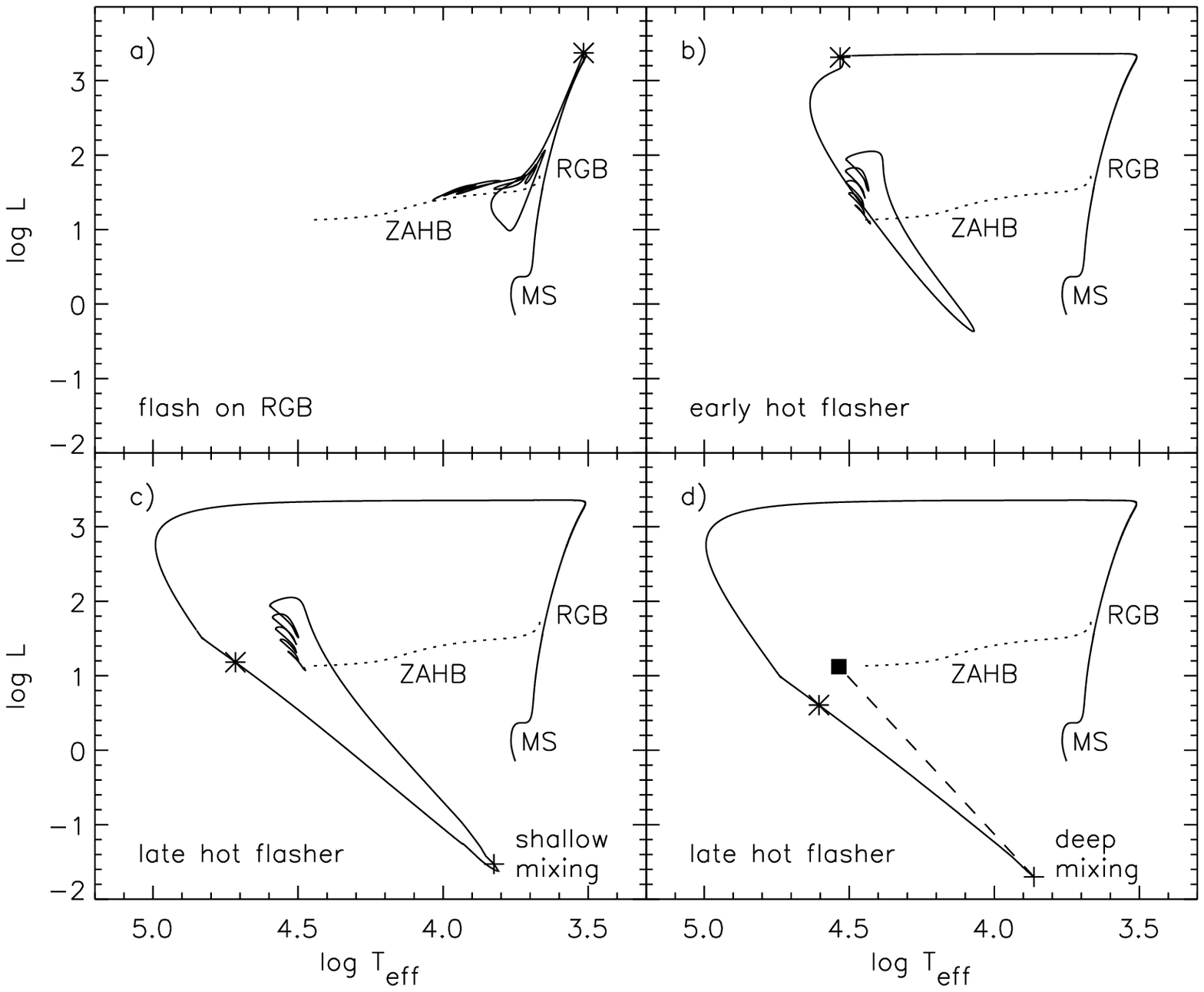}
\figurenum{2}
\caption{Evolution of a solar metallicity star from the main sequence (MS)
through the helium flash to the ZAHB (dotted curve) for
different amounts of mass loss on the red giant branch. For
sufficiently
large mass loss, a star evolves off the RGB to high effective
temperatures
before igniting helium as either an early or late hot flasher. The
peak of the helium flash is indicated by an asterisk. The flash
convection zone reached the hydrogen envelope at the plus sign
along the tracks in panels {\it c\/} and {\it d\/}. These panels
illustrate the two types of flash mixing: shallow mixing in which
the hydrogen envelope is mixed only with the convective shell in the
outer part of the core and deep mixing in which the hydrogen envelope
is mixed all the way into the site of the flash. The
model calculations in panel {\it d\/} were stopped at the onset
of deep mixing, and a ZAHB model (solid square) was then computed
assuming a helium- and carbon-rich envelope composition.  The
evolution during this phase is shown schematically by the dashed
line.  Flash mixing did not occur for the canonical sequences in
panels {\it a\/} and {\it b\/}.}
\end{figure}
\clearpage

\begin{figure}
\epsscale{0.75}
\plotone{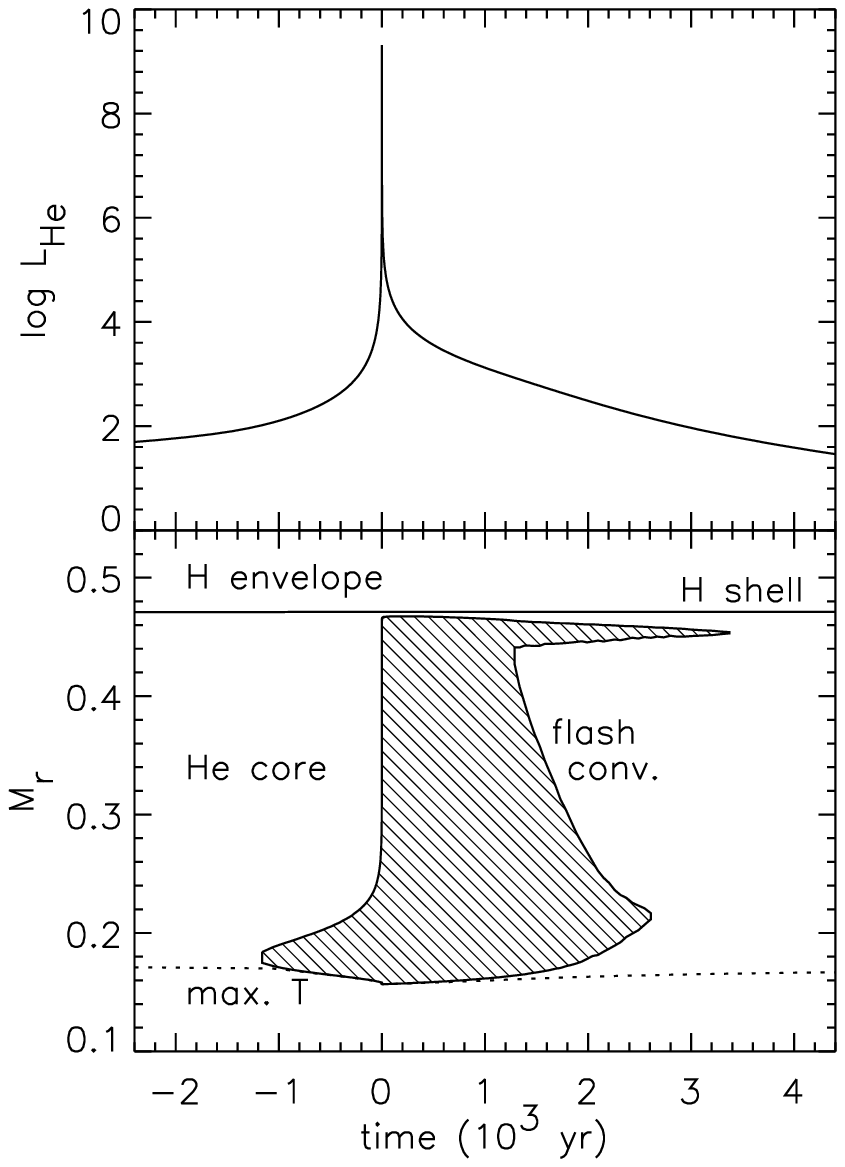}
\figurenum{3}
\caption{Time dependence of the helium-burning luminosity $\log L_{He}$
(upper panel) and the mass coordinate $M_r / M_\odot$ at the edge of the
flash convection zone (lower panel, shaded region) during a canonical
helium flash. The zero-point of the timescale corresponds to the
peak of the flash. The location of the hydrogen shell is also given
in the lower panel. The helium flash occurs off-center at the point
where the temperature in the core is a maximum (dotted curve). Note
that a convective shell persists in the outer part of the core
after the main body of the flash convection zone recedes.  During a
canonical flash, the flash convection zone does not reach the
hydrogen envelope, and the surface composition is unchanged.}
\end{figure}
\clearpage

\begin{figure}
\epsscale{0.85}
\plotone{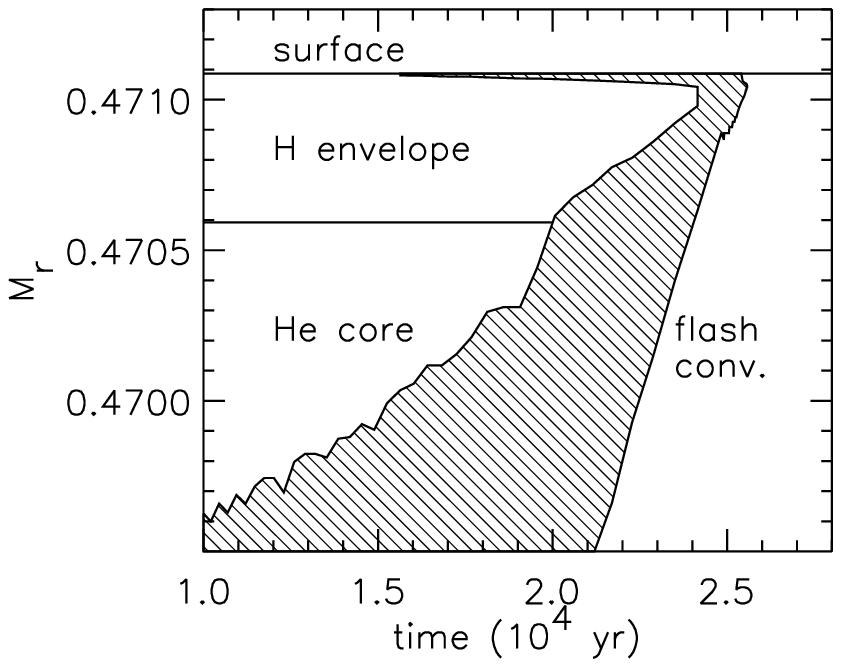}
\figurenum{4}
\caption{Time dependence of the mass coordinate $M_r / M_\odot$ at the
edge of the convective shell in the outer part of the core
during the phase of shallow mixing for the sequence
plotted in Fig. 2c. The horizontal line at $M_r = 0.4706\,M_\odot$ gives
the location of the hydrogen shell. The zero point of the timescale
corresponds to the peak of the helium flash. The fluctuations in the
edges
of the convective shell are due to the finite zoning of the mesh points
in the models. The convective shell
reaches the hydrogen envelope at time t = $2\times10^4$ yr. Helium
and carbon from the core are then transported
into the envelope and eventually to the surface at t = $2.4\times10^4$\,yr.}
\end{figure}
\clearpage

\begin{figure}
\plotone{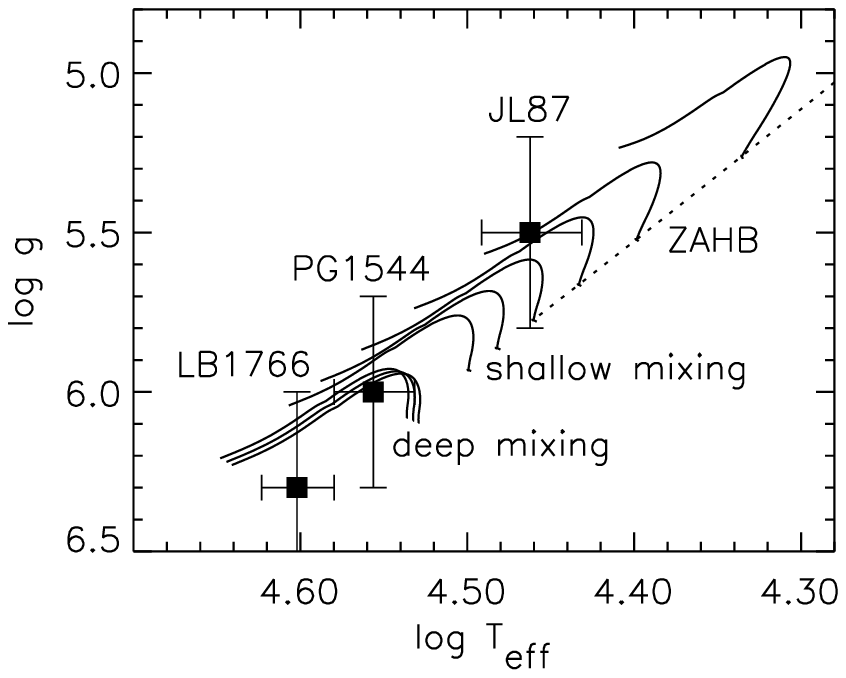}
\figurenum{5}
\caption{Comparison of the stellar parameters for the He-rich sdB
stars, PG1544+488,
JL87, and LB1766, with EHB evolutionary tracks for canonical models
(reddest 4 tracks),
models with shallow mixing (intermediate 2 tracks), and models with
deep mixing
(bluest 3 tracks). The dotted line denotes the canonical ZAHB.
\label{HRDfig}}
\end{figure}
\clearpage

\begin{figure}
\plotone{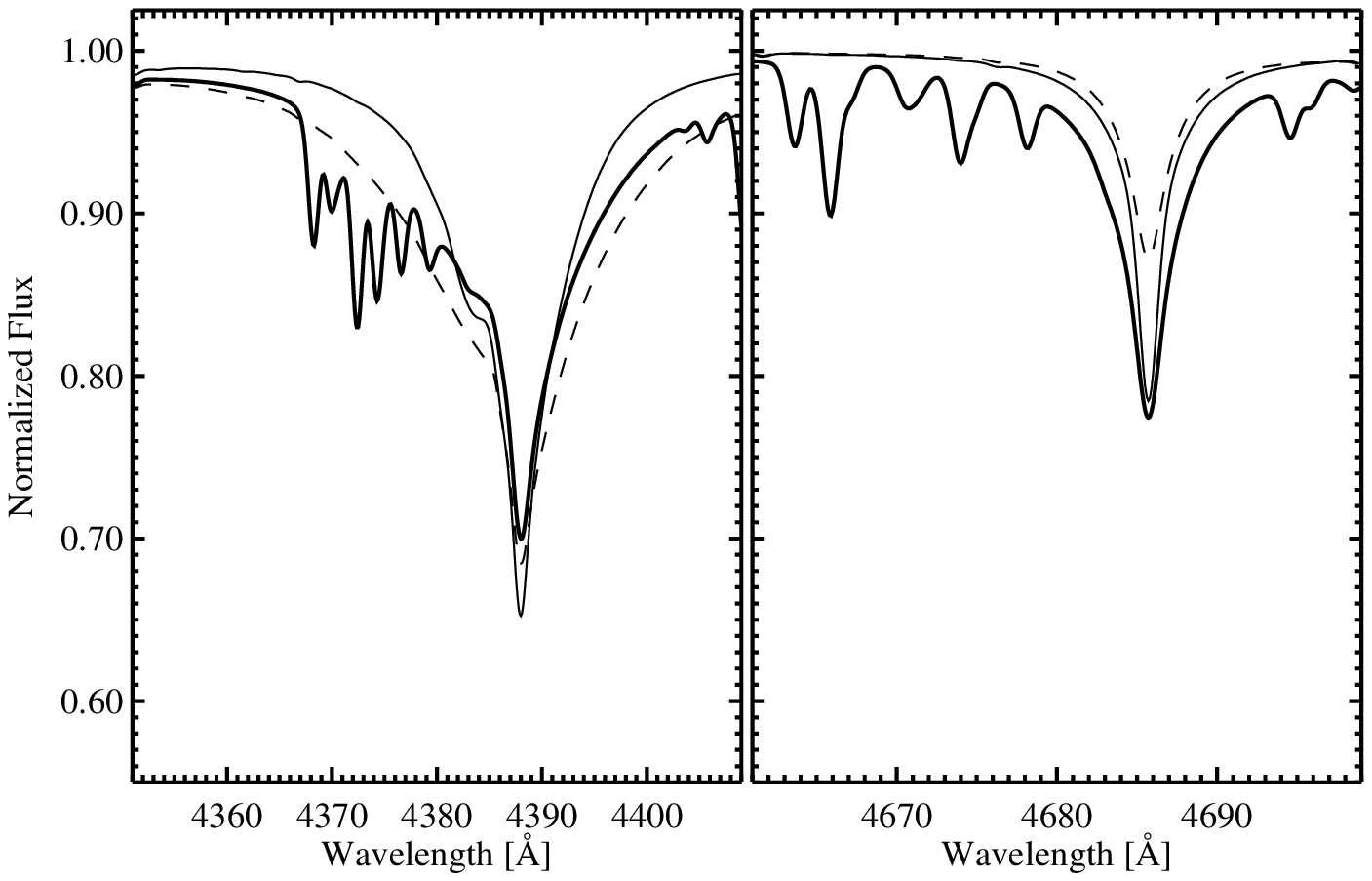}
\figurenum{6}
\caption{\ion{He}{1}\,$\lambda$4388 and \ion{He}{2}\,$\lambda$4686 line
profiles predicted by NLTE model atmospheres (\teff = 36,000\,K). Line
profiles for H-He model atmospheres, $Y=0.998$, $\log g=5.0$ (full thin  
line)
and $\log g=6.0$ (dashed thin line), are compared to the line profiles
calculated with the C-rich reference model atmosphere (full thick  
line). \label{BlueFig}}
\end{figure}


\clearpage
\begin{deluxetable}{lllrllr}
\tablecaption{\fuse\  observing log.  \label{ObsTbl}}
\tablehead{ \colhead{Star} & \colhead{$\alpha$(J2000)} & \colhead{$\delta$(J2000)} &
    \colhead{$B$} & \colhead{${\rm FUSE}ID$} & \colhead{Date} & \colhead{Exp. time [ksec]} }
\startdata
   PG1544+488 & 15 46 11.7 & +48 38 37 & 12.8    & C1290101 & 2002-03-26 & 7.270 \\
   \nodata    & \nodata    &  \nodata  & \nodata & C1290102 & 2002-07-14 & 8.541 \\
   JL87       & 21 48 37.9 & -76 20 45 & 12.0    & C1290401 & 2002-06-10 & 12.863 \\
   LB1766     & 04 59 19.3 & -53 52 55 & 12.3    & C1290201 & 2003-01-03 & 5.353 \\
\enddata
\end{deluxetable}


\clearpage
\begin{deluxetable}{llll}
\tablecaption{Results from the analysis of the \fuse\  spectra
\label{ResuTbl}}
\tablehead{ \colhead{Star} & \colhead{PG1544+488} & \colhead{JL87} &
    \colhead{LB1766} }
\startdata
{\bf Stellar Parameters} \\
\teff\ [K]              &    36000     &    29000    &    40000 \\
$\log g$ [cm/s$^2$]     &      6.0     &      5.5    &      6.3 \\
$\log L$ [$L_\odot$] \tablenotemark{a}    &      1.3     &      1.4 &      1.2 \\
$V \sin i$ [km/s]       &      100      &      30    &   $\leq$25  \\[2mm]
{\bf Surface Abundances} (mass fraction) \\
H                       & $<0.002$ &  $0.55 - 0.70$   &  $<0.0025$ \\
He                      & 0.96&  $0.43 - 0.28$       &   0.99   \\
C                       &  0.02       &  0.014       &  0.0001  \\
N                       &  0.01       &  0.004       &  0.006 \\
O                       &  \nodata     &  \nodata    &  0.003 \\
Si                      & $6\times 10^{-4}$ & $7\times 10^{-4}$& $3\times 10^{-4}$ \\
P                       & $2\times 10^{-5}$      & \nodata&  \nodata  \\
S                       & $3\times 10^{-4}$      & \nodata&  \nodata  \\
Fe                      & 0.0013       & 0.0013      &  0.0013   \\[2mm]
{\bf Interstellar Column Densities} [cm$^{-2}$] \\
\ion{H}{1}              & $3\times 10^{18}$ & $3\times 10^{20}$& $<1\times 10^{18}$ \\
H$_2$                   & $<5\times 10^{14}$ & $1\times 10^{20}$& $<5\times 10^{14}$ \\
\enddata
\tablenotetext{a}{Calculated, assuming a stellar mass, $M=0.5\,M_\odot$.}
\end{deluxetable}

\end{document}